\documentclass[aps,reprint,twocolumn,superscriptaddress,citeautoscript,showpacs,amsmath,amssymb,floatfix,prb]{revtex4-1}

\usepackage{amsmath}
\usepackage{amssymb}
\usepackage{graphicx}
\usepackage{times}
\usepackage{bm}
\usepackage{color}
\usepackage{gensymb}
\usepackage{dcolumn}

\newcolumntype{d}[1]{D{.}{.}{#1}}

\begin{document}

\title{Reduction of the ordered magnetic moment and its relationship to Kondo coherence in Ce$_{1-x}$La$_{x}$Cu$_{2}$Ge$_{2}$}

\author{B. G. Ueland}
\affiliation{Ames Laboratory, U.S. DOE, Iowa State University, Ames, Iowa 50011, USA}
\affiliation{Department of Physics and Astronomy, Iowa State University, Ames, Iowa 50011, USA}

\author{N. H. Jo}
\affiliation{Ames Laboratory, U.S. DOE, Iowa State University, Ames, Iowa 50011, USA}
\affiliation{Department of Physics and Astronomy, Iowa State University, Ames, Iowa 50011, USA}

\author{A. Sapkota}
\affiliation{Ames Laboratory, U.S. DOE, Iowa State University, Ames, Iowa 50011, USA}
\affiliation{Department of Physics and Astronomy, Iowa State University, Ames, Iowa 50011, USA}

\author{W. Tian}
\affiliation{Quantum Condensed Matter Division, Oak Ridge National Laboratory, Oak Ridge, Tennessee 37831, USA}

\author{M. Masters}
\affiliation{Ames Laboratory, U.S. DOE, Iowa State University, Ames, Iowa 50011, USA}
\affiliation{Department of Physics and Astronomy, Iowa State University, Ames, Iowa 50011, USA}

\author{H. Hodovanets}
\altaffiliation[Present Address:  ]{Department of Physics, University of Maryland, College Park, Maryland 20742, USA}
\affiliation{Ames Laboratory, U.S. DOE, Iowa State University, Ames, Iowa 50011, USA}
\affiliation{Department of Physics and Astronomy, Iowa State University, Ames, Iowa 50011, USA}

\author{S. S. Downing}
\affiliation{Ames Laboratory, U.S. DOE, Iowa State University, Ames, Iowa 50011, USA}
\affiliation{Department of Physics and Astronomy, Iowa State University, Ames, Iowa 50011, USA}

\author{C. Schmidt}
\affiliation{Ames Laboratory, U.S. DOE, Iowa State University, Ames, Iowa 50011, USA}
\affiliation{Department of Physics and Astronomy, Iowa State University, Ames, Iowa 50011, USA}

\author{R. J. McQueeney}
\affiliation{Ames Laboratory, U.S. DOE, Iowa State University, Ames, Iowa 50011, USA}
\affiliation{Department of Physics and Astronomy, Iowa State University, Ames, Iowa 50011, USA}

\author{S. L. Bud'ko}
\affiliation{Ames Laboratory, U.S. DOE, Iowa State University, Ames, Iowa 50011, USA}
\affiliation{Department of Physics and Astronomy, Iowa State University, Ames, Iowa 50011, USA}

\author{A. Kreyssig}
\affiliation{Ames Laboratory, U.S. DOE, Iowa State University, Ames, Iowa 50011, USA}
\affiliation{Department of Physics and Astronomy, Iowa State University, Ames, Iowa 50011, USA}

\author{P. C. Canfield}
\affiliation{Ames Laboratory, U.S. DOE, Iowa State University, Ames, Iowa 50011, USA}
\affiliation{Department of Physics and Astronomy, Iowa State University, Ames, Iowa 50011, USA}

\author{A. I. Goldman}
\affiliation{Ames Laboratory, U.S. DOE, Iowa State University, Ames, Iowa 50011, USA}
\affiliation{Department of Physics and Astronomy, Iowa State University, Ames, Iowa 50011, USA}

\date{\today}

\begin{abstract}
The microscopic details of the suppression of antiferromagnetic order in the Kondo-lattice series Ce$_{1-x}$La$_{x}$Cu$_{2}$Ge$_{2}$ due to nonmagnetic dilution by La are revealed through neutron diffraction results for $x=0.20$, $0.40$, $0.75$, and $0.85$.  Magnetic Bragg peaks are found for $0.20\le x\le0.75$, and both the N\'{e}el temperature, $T_{\textrm{N}}$, and the ordered magnetic moment per Ce, $\mu$, linearly decrease with increasing $x$.  The reduction in $\mu$ points to strong hybridization of the increasingly diluted Ce $4f$ electrons, and we find a remarkable quadratic dependence of $\mu$ on the Kondo-coherence temperature.  We discuss our results in terms of local-moment- versus itinerant-type magnetism and mean-field theory, and show that Ce$_{1-x}$La$_{x}$Cu$_{2}$Ge$_{2}$ provides an exceptional opportunity to quantitatively study the multiple magnetic interactions in a Kondo lattice.
\end{abstract}

\maketitle
Kondo lattices consist of a periodic arrangement of localized magnetic moments (e.g.~$4f$ spins) coupled to itinerant charge carriers through the Kondo interaction \cite{Doniach_1977, Stewart_1984, Brandt_1984, Stewart_2001, Stewart_2006, Maple_2005, Gulacsi_2006, Coqblin_2006}.  The Kondo coupling tends to screen the localized spins \cite{Kondo_1964} and competes with the conduction-electron-mediated Ruderman-Kittel-Kasuya-Yosida (RKKY) exchange interaction between them. Tuning the relative strength of the intra-site Kondo to inter-site RKKY exchange interaction \cite{Doniach_1977, Coqblin_2006} leads to various phenomena including long-range magnetic order, unconventional superconductivity, quantum-critical fluctuations, and heavy-Fermi- and non-Fermi-liquid behaviors \cite{Stewart_1984, Stewart_2001, Stewart_2006, Maple_2005, Si_2013}.  

In $4f$ Kondo-lattice compounds, heavy-fermion behavior is characterized by an enhanced effective mass of itinerant charge carriers due to band hybridization between localized $4f$ and itinerant electrons.  In many heavy-fermion metals, a peak or change in slope of the resistivity occurs upon cooling through a temperature $T_{\textrm{coh}}$, which signals a crossover from incoherent to coherent scattering and the onset of Kondo coherence \cite{Coleman_2007}.  An overall understanding of Kondo coherence and its relationship to magnetic order remains elusive.

A recent study on the tetragonal Kondo-lattice series Ce$_{1-x}$La$_{x}$Cu$_{2}$Ge$_{2}$, in which the Ce magnetic sublattice is diluted with nonmagnetic La, remarkably found that AFM order and Kondo coherence coexist up to $x=0.80$, and that Kondo coherence persists up to $x=0.90$  \cite{Hodovanets_2015}.  This implies a small percolation limit for the magnetic sublattice ($\approx9\%$), and is consistent with a $3$-dimensional network possessing further-than-nearest-neighbor magnetic-exchange interactions \cite{Hodovanets_2015}.  The robustness of the AFM order and its coexistence with Kondo coherence at high dilution levels calls for a microscopic examination of the evolution of the AFM order as $x$ is increased.

Here, we present results from neutron diffraction experiments on the series that elucidate the changes to the ordered magnetic moment per Ce $\mu$, the antiferromagnetic propagation vector $\bm{\tau}$, and the N\'{e}el temperature $T_{\textrm{N}}$ caused by nonmagnetic dilution.  We find that both $T_{\textrm{N}}$ and  $\mu$ linearly decrease, and that the $l$ component of $\bm{\tau}$ increases with increasing $x$.  Surprisingly, we determine that $\mu$ is quadratically related to $T_{\textrm{coh}}$ for at least $x\le0.75$, which provides a microscopic relationship between the size of the ordered moment and Kondo coherence in these compounds.

CeCu$_{2}$Ge$_{2}$ has the same tetragonal ThCr$_{2}$Si$_{2}$ crystal structure (space group $I4/mmm$)  as the heavy-fermion superconductor CeCu$_{2}$Si$_{2}$ \cite{Stewart_1984,Krimmel_1997}.   A maximum in its resistivity at $T_{\textrm{coh}}\approx5.5$~K signals the onset of Kondo coherence, and heat capacity $C_{\textrm{p}}$ data show a moderate effective-mass enhancement characterized by $C_{\textrm{p}}/T=0.245$~J$/$mol-K$^{2}$ at $T=0.39$~K  \cite{Hodovanets_2015}.  Complex AFM order appears below $T_{\textrm{N}} = 4.2$~K \cite{Knopp_1989,Loidl_1992a,Krimmel_1997}, consisting of a modulated structure with $\bm{\tau}= (0.284(1), 0.284(1), 0.543(1))$ and  $\mu=1.1(1)~\mu_{\textrm{B}}/$Ce, at $T = 1.5$~K, lying in the \textbf{ab} plane. 

Single-crystals of Ce$_{1-x}$La$_{x}$Cu$_{2}$Ge$_{2}$ were grown from Cu-Ge self-flux as previously described \cite{Hodovanets_2015}. Samples with nominal values of $x$ of $0.20$, $0.40$, $0.75$, and $0.85$, and masses $m=0.59$, $1.54$, $1.86$, and $1.04$~g, respectively, were selected for the diffraction experiments.  $C_{\textrm{p}}(T)$ for $x=0.85$ was measured down to $T=0.06$~K using the dilution refrigerator option of a Quantum Design Physical Property Measurement System and the semi-adiabatic heat pulse technique. Neutron diffraction data were recorded using the HB-$1$A Fixed-Incident-Energy Triple-Axis Spectrometer at the High Flux Isotope Reactor, Oak Ridge National Laboratory. Incident- and final-neutron energies of $E=14.6$~meV were used, and effective collimations of $40^{\prime}$-$40^{\prime}$-$40^{\prime}$-$80^{\prime}$ existed before the pyrolytic graphite (PG) $(0~0~2)$ monochromator, between the monochromator and sample, the sample and PG $(0~0~2)$ analyzer, and the analyzer and detector, respectively.  Two PG filters were placed in the incident beam to suppress higher-order harmonics.  Samples were mounted with their $(h~h~l)$ reciprocal-lattice planes coincident with the scattering plane, and were cooled down using either an orange-type cryostat ($x=0.20$), a $^{3}$He insert ($x=0.40$ and $0.75$), or a dilution-refrigerator insert ($x=0.85$).  The base temperatures were $T=1.6$, $0.3$, and $0.04$~K, respectively. Momentum transfers are expressed in reciprocal-lattice units (r.l.u.).  Lattice parameters determined from the neutron diffraction data are given in the Supplemental Materials \cite{SM} and compare well to previous results \cite{Hodovanets_2015}. 

\begin{figure}[]
\centering
\includegraphics[width=1\linewidth]{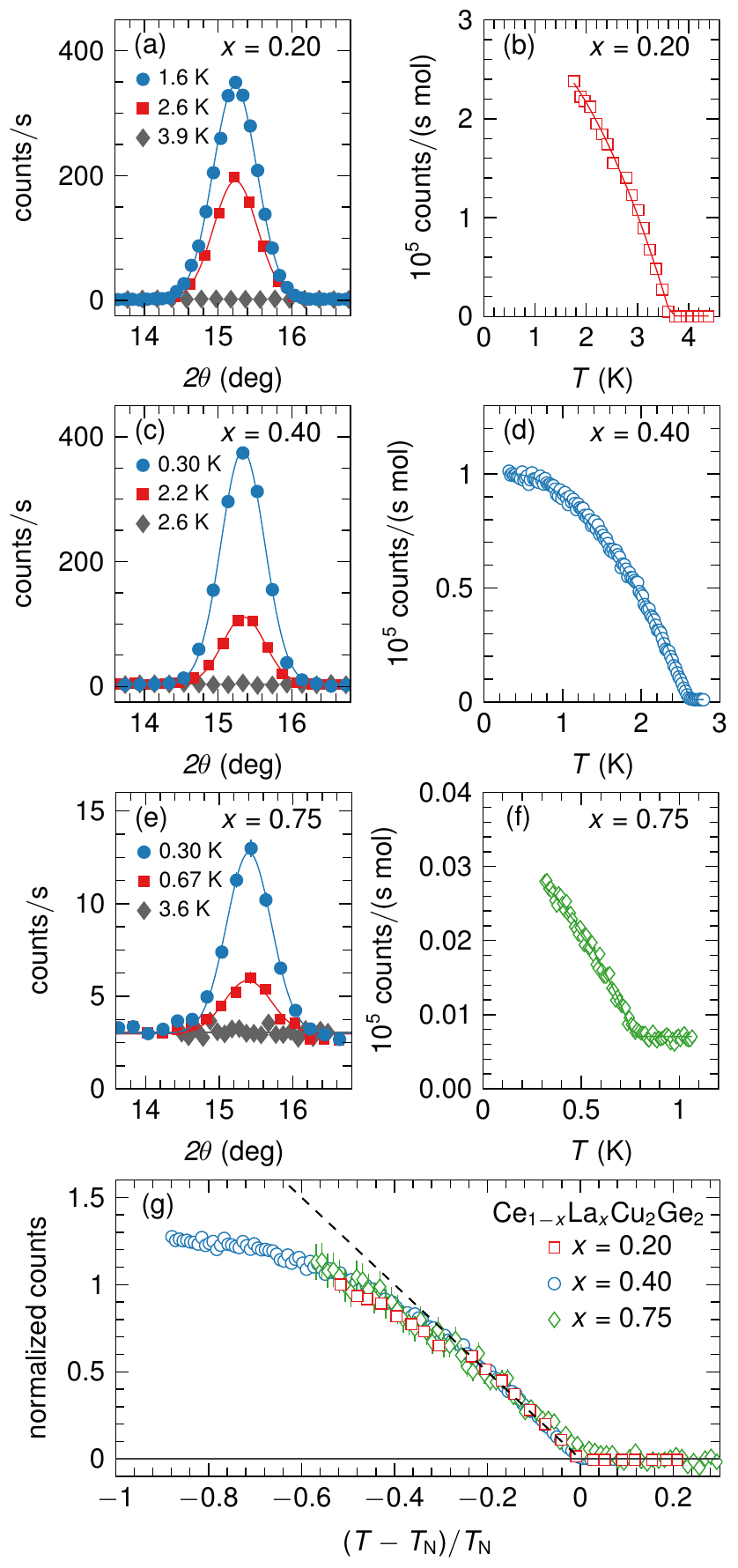}
\caption{Data from $\theta$-$2\theta$ (longitudinal) scans through the $(0~0~0)+\tau$ magnetic Bragg peak at different temperatures and the peak's intensity versus temperature for $x=0.20$ [(a) and (b)], $0.40$ [(c) and (d)], and $0.75$ [(e) and (f)].  The peaks are fit by Gaussian lineshapes. Data in (b), (d), and (f) are normalized per mole formula unit, and are replotted in (g) versus reduced temperature after being normalized to $1$ at $\frac{T-T_{\textrm{N}}}{T_{\textrm{N}}}=-0.5$ and $0$ at $T>T_{\textrm{N}}$.  The dashed line corresponds to $\beta=0.5$.  The finite intensity above $T_{\textrm{N}}$, obvious in (f), is due to incoherent scattering and fast-background counts.}
\label{Fig1}
\end{figure}

Data from $\theta$-$2\theta$ (longitudinal) scans through the $(0~0~0)+\tau$ magnetic Bragg peak at various temperatures are shown in Figs.~\ref{Fig1}(a),\ref{Fig1}(c), and\ref{Fig1}(e) for $x=0.20$, $0.40$, and $0.75$, respectively.  The peaks are fit to Gaussian lineshapes, with full-widths-at-half maximum of $\approx43^{\prime}$, which is comparable to the tightest effective collimation used.  Though not shown, data from rocking scans also show a single peak for each $x$, with a width corresponding to the tightest collimation used. Magnetic Bragg peaks were also found for $(1~1~0)\pm\tau$, $(0~0~2)+\tau$, and $(1~1~2)-\tau$.  Data for $x=0.85$ are shown below.  

Figures~\ref{Fig1}(b), \ref{Fig1}(d), and \ref{Fig1}(f) show the intensity of the $(0~0~0)+\tau$ magnetic Bragg peak as a function of temperature for $x=0.20$, $0.40$ and $0.75$.  The plots give $T_{\textrm{N}}=3.6(2)$, $2.6(1)$, and $0.80(7)$~K, respectively. Since the scattering intensity depends on the square of the ordered magnetic moment, these data show that the ordered moment per mole formula unit decreases with increasing $x$.  Figure~\ref{Fig1}(g) gives the normalized scattering intensities versus reduced temperature, where the dashed line indicates a magnetic critical exponent of $\beta=0.5$, which corresponds to mean-field type behavior.

\begin{figure}[]
\centering
\includegraphics[width=1.0\linewidth]{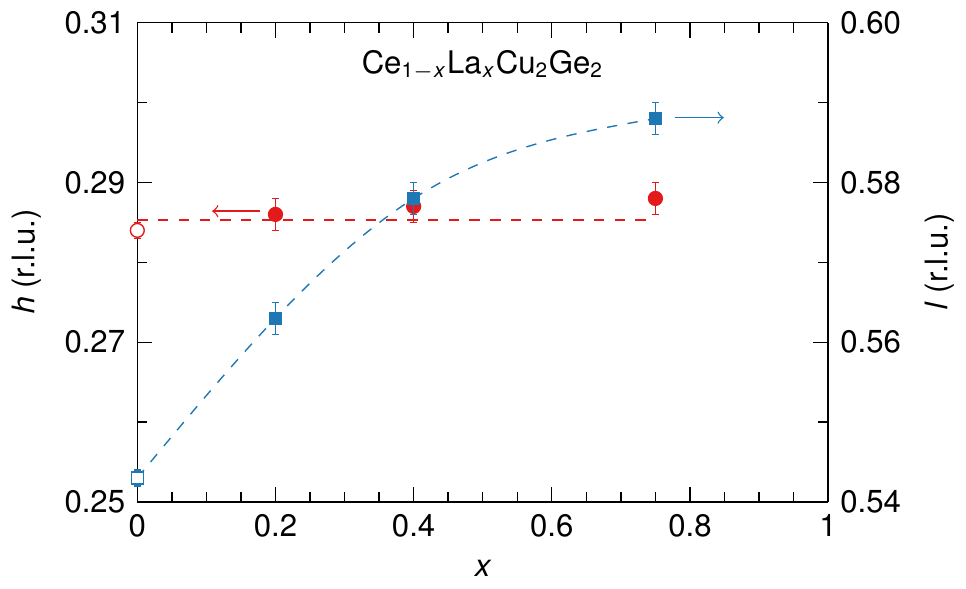}
\caption{The components of $\bm{\tau}=(h, h, l)$ versus $x$ in reciprocal lattice units (r.l.u.). Values for $x=0$ are from Ref.~\onlinecite{Krimmel_1997}.  The straight line is a fit to the data, and the curved line is a guide to the eye.}
\label{Fig2}
\end{figure}

 Longitudinal and rocking scans made through the $(0~0~0)+\tau$, $(1~1~0)-\tau$, $(0~0~2)+\tau$, and $(1~1~2)-\tau$ magnetic Bragg peaks at base temperature were used to determine $\bm{\tau}$, the components of which are given versus $x$ in Fig.~\ref{Fig2}.  $h(x)$ is practically constant whereas $l(x)$ monotonically increases by $8.3\%$ between $x=0$ and $0.75$.  $\bm{\tau}$ is not temperature dependent for any of our samples, however, the reported change in $\bm{\tau}(x=0)$ with temperature \cite{Krimmel_1997} is near the resolution limit of our experiment.  $l(x)$ is not constant and increases with $x$, which shows a decoupling of changes to $\bm\tau$ and the unit-cell parameters \cite{SM}.

\begin{figure}[]
\centering
\includegraphics[width=1.0\linewidth]{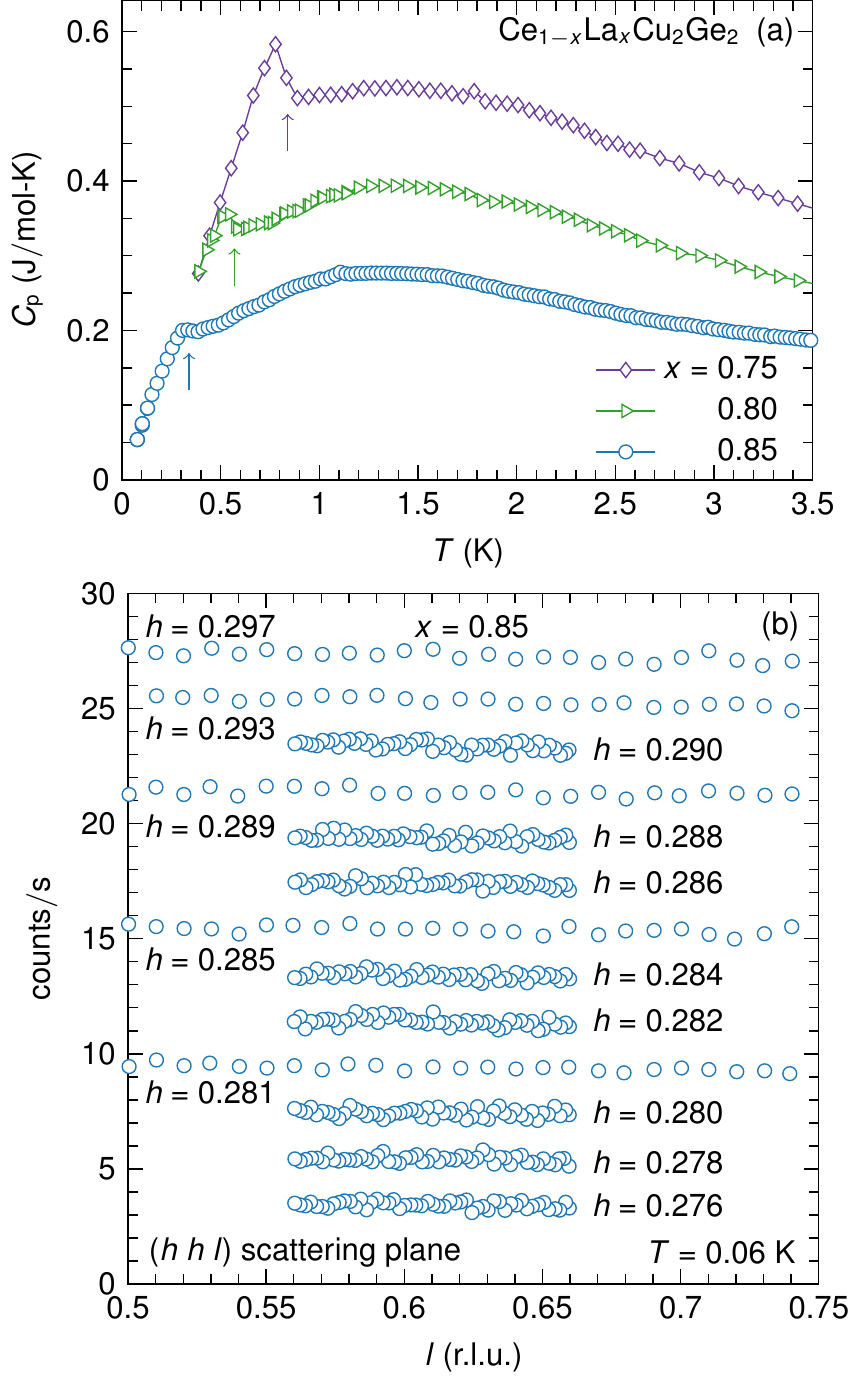}
\caption{(a) Heat capacity data for $x=0.75$, $0.80$, and $0.85$.  Data for $x=0.75$ and $0.80$ are also shown in Ref.~\onlinecite{Hodovanets_2015}.  Arrows mark peaks associated with the AFM transition.  (b) Neutron diffraction data for $x=0.85$ from scans along $l$ performed at $T=0.06$~K for the values of $h$ indicated.  Data for values of $h$ ending with an even number are for a beam monitor value corresponding to $2.2$ minutes$/$point, whereas data for $h$ ending with an odd number correspond to $3$ minutes$/$point.  Error bars are within the symbol size, and datasets are offset by $2$~counts$/$s.}
\label{Fig3}
\end{figure}

Heat-capacity data for $x=0.85$ are plotted in Fig.~\ref{Fig3}(a), along with data for $x=0.75$ and $0.80$ from Ref.~\onlinecite{Hodovanets_2015}.  For $x=0.75$ and $0.80$, peaks at $T_{\textrm{N}}=0.84$ and $0.57$~K mark the AFM transition, and the broad maxima at higher temperatures are associated with the Kondo temperature $T_{\textrm{K}}$.  Data for the $3$ samples are qualitatively similar, which suggests that the peak at $T_{\textrm{N}}=0.34(3)$~K for $x=0.85$ signals an AFM transition. Details concerning analysis of the heat capacity data, as well as the determination of $T_{\textrm{K}}$ and $T_{\textrm{coh}}$ are given in Ref.~\onlinecite{Hodovanets_2015}. The data in Fig.~\ref{Fig3}(a) are replotted in the Supplemental Material \cite{SM} as $[C/T](T)$ and normalized in units of J$/$mol$_{\textrm{Ce}}$-K$^{2}$.

Figure~\ref{Fig3}(b) shows neutron diffraction data for $x=0.85$ at $T=0.06$~K from scans performed across a region of reciprocal space where a magnetic Bragg peak corresponding to $(0~0~0)+\tau$ is expected based on Fig.~\ref{Fig2}. No such peak is seen despite the $C_{\textrm{p}}(T)$ data indicating $T_{\textrm{N}}=0.34(3)$~K.  Data from scans across the expected $(1~1~2)-\tau$ position at $T=0.06$~K also do not show a Bragg peak.  Two possibilities for the peak's absence are: ($1$) $\mu$ is smaller than the detection limit, and ($2$) $\bm{\tau}$ is much different than found for the other compositions. Regarding point ($1$), we estimate from the data and structure factors for $x=0.75$ that the smallest ordered moment detectable by the experiment is $\mu\approx0.06~\mu_{\textrm{B}}/$Ce.  This point is further discussed below.

With respect to point ($2$), data from line scans at $T=0.06$~K across the $(1~1~2)$ structural Bragg peak along the $(h,h,0)$ and $(0,0,l)$ high-symmetry directions, spanning $(0.5~0.5~2)$ to $(1.5~1.5~2)$ and $(1~1~0.5)$ to $(1~1~1.5)$, show no magnetic Bragg peaks.  This rules out,  within the sensitivity of our measurements, AFM order with a $\bm{\tau}$ corresponding to the investigated ranges. No magnetic Bragg peaks were detected at the $(0~0~3)$ and $(1~1~1)$ positions as well, and data from scans through the $(0~0~2)$ and $(1~1~0)$ positions at $T=0.06$ and $0.7$~K reveal no changes suggesting the onset of ferromagnetic order with decreasing temperature.  Nevertheless, these measurements do not completely rule out the presence of weak ferromagnetism nor a dramatic change to the magnetic structure.

Two models have been employed to describe the AFM order for $x=0$: a spin-density wave with a sinusoidally-modulated amplitude (SDW) \cite{Krimmel_1997} and a spiral \cite{Knopp_1989, Singh_2011}.  Both have $\mu$ lying in the \textbf{ab} plane.  We determined $\mu$ associated with each model using the integrated intensities of the $(0~0~0)+\tau$, $(1~1~0)-\tau$, $(0~0~2)+\tau,$ and $(1~1~2)-\tau$ magnetic Bragg peaks, and the structure factors and integrated intensities of the $(0~0~2)$, $(1~1~0)$, and $(0~0~4)$ structural Bragg peaks for $x=0.20$, $0.40$, and $0.75$.  Both AFM models have similar structure factors, and the presumed presence of magnetic domains does not allow for distinguishing between the two with our data.  Nevertheless, the ratio of $\mu$ for the spiral model to that for the SDW is simply $\sqrt{2}$.

\begin{figure}
 \centering
 \includegraphics[width=1.0\linewidth]{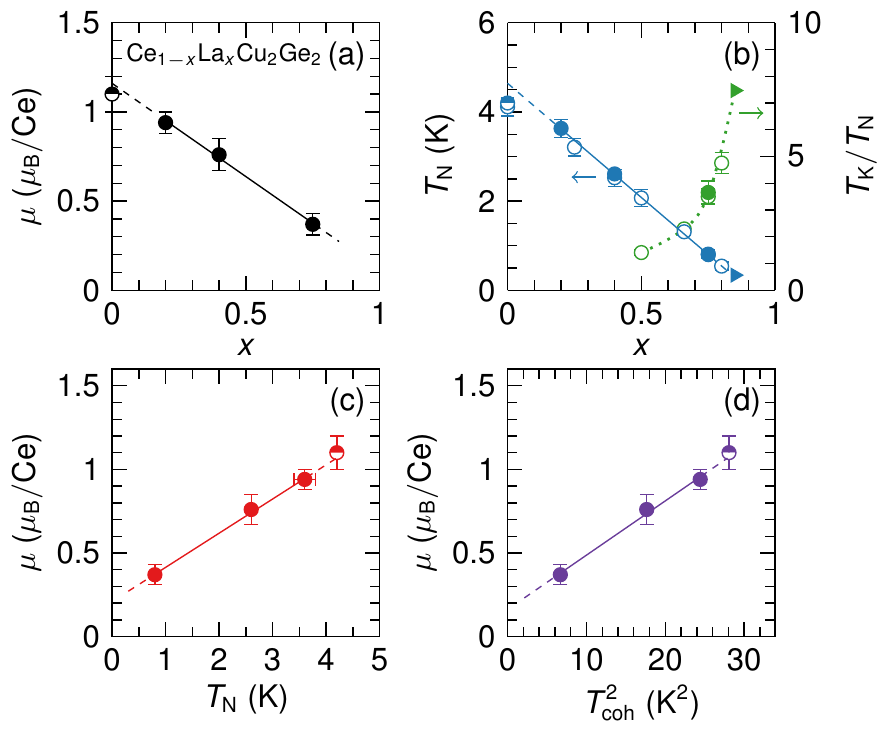}
 \caption{(a) $\mu(x)$, (b) $T_{\textrm{N}}(x)$ (left axis), $[T_{\textrm{K}}/T_{\textrm{N}}](x)$ (right axis), (c) $\mu(T_{\textrm{N}})$, and (d) $\mu(T_{\textrm{coh}}^{2})$.  Values for $\mu$ are based on the SDW model, and all $4$ values are for a reduced temperature of $\frac{T-T_{\textrm{N}}}{T_{\textrm{N}}}=-0.89$. Filled circles indicate our neutron diffraction results, and the triangles in (b) mark points using $T_{\textrm{N}}(x=0.85)$ from our heat capacity measurements.  Data from Ref.~\onlinecite{Krimmel_1997} (half-filled circles) and Fig.\,$4$ of Ref.~\onlinecite{Hodovanets_2015} (open circles) are included.  The values for $T_{\textrm{coh}}^{2}$ and $T_{\textrm{K}}$ are from Ref.~\onlinecite{Hodovanets_2015}.  Solid lines are fits to data from our neutron diffraction experiments, whereas dashed lines indicate extrapolations to $x=0$ and $0.85$.  The dotted line in (b) is a guide to the eye.}
 \label{Fig4}
 \end{figure}

The values for $\mu$ using the SDW model, after correcting for the expected number of magnetic domains, are given in Figs~\ref{Fig4}.  All of the plotted values for $\mu$ are for a reduced temperature of $\frac{T-T_{\textrm{N}}}{T_{\textrm{N}}}=-0.89$.  Since Fig.~\ref{Fig1}(g) shows that the scattering intensity versus temperature plots for the magnetic Bragg peak follow similar behavior for $x=0.20$, $0.40$. and $0.75$, the values for $x=0.20$ and $0.75$ were extrapolated down to $\frac{T-T_{\textrm{N}}}{T_{\textrm{N}}}=-0.89$ by assuming that they follow the same temperature dependence as $x=0.40$.  This is done in order to obtain values of $\mu$ at a low reduced temperature.  The value of $\mu$ for $x=0$ was obtained from data in Ref.~\onlinecite{Krimmel_1997}.
 
Figures~\ref{Fig4}(a) and \ref{Fig4}(b) reveal that both $\mu(x)$ and $T_{\textrm{N}}(x)$ linearly decrease with increasing $x$.  The simple $x$ dependence of both quantities contrasts with, for example, data for the well known quantum-critical heavy-fermion series CeCu$_{6-x}$Au$_{x}$ showing that $T_{\textrm{N}}$ decrease linearly with $x$ but $\mu$ does not \cite{Lohneysen_1998, Schroder_2000}.  The line determined from fitting $T_{\textrm{N}}(x)$ agrees with previous results \cite{Hodovanets_2015}, and the extrapolation of the fit to $\mu(x)$ approaches the value of $\mu(x=0)$ determined at $\frac{T-T_{\textrm{N}}}{T_{\textrm{N}}}=-0.89$ from data given in Ref.~\onlinecite{Krimmel_1997}. Since earlier work found $T_{\textrm{N}}\sim T_{\textrm{coh}}^{2}$ \cite{Hodovanets_2015}, we determined $T_{\textrm{coh}}$ using Fig.~$5$(b) in Ref.~\onlinecite{Hodovanets_2015} and the values of $T_{\textrm{N}}$ from our diffraction data.  $\mu(T_{\textrm{coh}}^{2})$ is plotted in Fig.~\ref{Fig4}(d), which, remarkably, shows that $\mu\sim T^{2}_{\textrm{coh}}$.

Noticeably, extrapolating the fit in Fig.~\ref{Fig4}(c) gives a finite value for $\mu$ as $T_{\textrm{N}}\rightarrow0$~K.  This is unexpected, since the Doniach description of a Kondo lattice predicts that Kondo coupling eventually dominates the RKKY exchange and suppresses AFM order \cite{Doniach_1977,Iglesis_1997}.  The green plot in Fig.~$4$(b), which incorporates data from Ref.~\onlinecite{Hodovanets_2015}, shows that $[T_{\textrm{K}}/T_{\textrm{N}}]$ increases rapidly at higher $x$, which may explain why magnetic Bragg peaks are not observed for $x=0.85$: the Kondo coupling is sufficiently strong to cause a much smaller $\mu$ than that predicted by the fit in Fig.~\ref{Fig4}(c).  In this scenario, $\mu$ must be below the experiment's detection limit of $\approx0.06~\mu_{\textrm{B}}/$Ce.  On the other hand, the Doniach description relies on changes to the Kondo-coupling strength and/or to the density of states at the Fermi-level, and it is not straightforward to tie $\mu(x)$ to either.

To try and understand the linear behavior of $T_{\textrm{N}}(x)$, one may begin from  mean-field theory for a bipartite lattice and treat the RKKY exchange as an effective Heisenberg interaction between localized spins \cite{Jensen_1991,Kittel_1996}.  This leads to \mbox{$T_\textrm{N}\sim z(1-x)J_{\textrm{RKKY}}\mu_{\textrm{eff}}^{2}$}, where $J_{\textrm{RKKY}}$ is the RKKY exchange strength, $z$ is the number of nearest-neighboring Ce in CeCu$_{2}$Ge$_{2}$, and $\mu_{\textrm{eff}}$ is the effective moment per Ce (as determined, for example, from magnetization measurements of the paramagnetic state) \cite{Kittel_1996}.  Here, a decrease in the number of localized spins linearly suppresses $T_{\textrm{N}}$ as long as $\mu_{\textrm{eff}}(x)$ is constant.  This happens, for example, in the site-diluted metallic Ising series Tb$_{1-x}$Y$_{x}$Ni$_{2}$Ge$_{2}$ \cite{Wiener_2000} and the magnetically-doped superconductor Y$_{x}$Gd$_{1-x}$Ni$_{2}$B$_{2}$C \cite{ElMassalami_1995}.

Turning to Kondo lattices, the RKKY exchange strength may be written as $J_{\textrm{RKKY}}\sim J^{2}_{\textrm{K}}\rho$, where $J_{\textrm{K}}$ is the Kondo-coupling strength, $\rho$ is the density of states at the Fermi level and $T_{\textrm{K}}\sim\exp(-1/J_{\textrm{K}}\rho)$ \cite{Doniach_1977, Coleman_2007}. Certain Kondo-lattice compounds have a characteristic temperature $T_{\textrm{ch}}$ (i.e.\ a temperature corresponding to a feature in thermodynamic or transport data) that decreases linearly over a range of increasing $x$.  $T_{\textrm{ch}}$ has been described by $T_{\textrm{ch}}(x)\sim J_{\textrm{RKKY}}(x)\sim J^{2}_{\textrm{K}}(x)(1-x)$ as long as $\mu_{\textrm{eff}}(x)$ is constant \cite{Brandt_1984}.  This relation is used to explain, for example, the change with $x$ of the spin-freezing temperature $T_{\textrm{sf}}$ for Ce$_{1-x}$La$_{x}$Cu$_{2}$Si$_{2}$ \cite{Brandt_1984}.  $T_{\textrm{sf}}(x)$ is linear for $x\approx0.7$ to $0.9$, but is not linear for lower $x$ due to $J_{\textrm{K}}$ depending on $x$.  A similar analysis is applied to $T_{\textrm{N}}(x)$ for Ce$_{1-x}$La$_{x}$Al$_{2}$Ga$_{2}$, which is linear for $x\le0.8$, and Ce$_{1-x}$Y$_{x}$Al$_{2}$Ga$_{2}$, for which a decrease in the unit-cell volume with increasing $x$ causes $J_{\textrm{K}}$ to increase and $T_{\textrm{N}}(x)$ to be nonlinear \cite{Garde_1991}.  Such analysis is also used for Ce$_{1-x}$(La,Y)$_{x}$(Au,Ag)$_{2}$Si$_{2}$ \cite{Garde_1994}.     

For Ce$_{1-x}$La$_{x}$Cu$_{2}$Ge$_{2}$, magnetization measurements over $T=150$ to $300$~K show that $\mu_{\textrm{eff}}(x)$ is constant \cite{Hodovanets_2015}, which suggests that the relations $T_{\textrm{N}}(x)\sim J_{\textrm{RKKY}}(x)\sim J^{2}_{\textrm{K}}(x)(1-x)$ may be applicable.  A relation between $J_{\textrm{K}}$ and $T_{\textrm{coh}}$ can be obtained from the two-fluid model for the Kondo-lattice, which treats $k_{\textrm{B}}T_{\textrm{K}}$ and $k_{\textrm{B}}T_{\textrm{coh}}$ as distinct energy scales  \cite{Nakatsuji_2004, Lonzarich_2012}, and subsequent work that links $J_{\textrm{K}}$ and $T_{\textrm{coh}}$ by assuming that $T_{\textrm{coh}}\propto J_{\textrm{RKKY}}\sim J^{2}_{\textrm{K}}\rho$ \cite{Yang_2008}.  This, however, leads to $T_{\textrm{N}}(x) \sim T_{\textrm{coh}}(x)$, which does not explain the quadratic dependence of $T_{\textrm{N}}$ on $T_{\textrm{coh}}$, nor $\mu\sim T_{\textrm{coh}}^{2}$.  Hence, these standard models for a Kondo-lattice, which attribute the existence of long-range magnetic order to localized spins interacting via RKKY magnetic exchange, do not readily explain the suppression of $\mu$ with increasing $x$.  This suggests stronger than expected hybridization of the Ce $4f$ electrons and the conduction band(s). 

In summary, the Kondo-lattice series Ce$_{1-x}$La$_{x}$Cu$_{2}$Ge$_{2}$ provides an exceptional opportunity for studying the boundary between localized and itinerant magnetism, as $\mu$ and the characteristic magnetic energy scales are experimentally determinable up to very high levels of nonmagnetic dilution, and the relations $\mu\sim T_{\textrm{N}}\sim T_{\textrm{coh}}^{2}$ microscopically link the magnetic order and Kondo coherence.  The small and decreasing value of $\mu$ with increasing $x$ presumably depends on changes to the hybridization of the localized Ce $4f$ and conduction electrons, and the small percolation limit of the magnetic sublattice implies that interactions out to the third nearest-neighbor are relevant \cite{Domb_1966,Kurzawski_2012}.  This suggests that mean-field theory should describe the magnetism, as has been done for other Kondo-lattice compounds \cite{Coqblin_2006,Coleman_2007,Iglesis_1997,Grenzebach_2006,Bernhard_2015}, however, we discuss above that standard considerations do not fully explain our results.  More generally, establishing further microscopic connections between long-range magnetic order and heavy-fermion behavior will lend insight into electronic properties arising from hybridization of localized and itinerant spins, as well as highlighting the differences between localized and itinerant magnetism.
    
\begin{acknowledgments}
We appreciate assistance from L.\ J.\ Santodonato while using the CG-$1$B utility diffractometer at HFIR to initially align the samples, and the Sample Environment Group at HFIR.  Work at the Ames Laboratory was supported by the Department of Energy, Basic Energy Sciences, Division of Materials Sciences \& Engineering, under Contract No.\ DE-AC02-07CH11358.  N.\ H.\ Jo is supported by the Gordon and Betty Moore Foundation EPiQS Initiative (Grant No.\ GBMF$4411$).  A portion of this research used resources at the High Flux Isotope Reactor, a U.\ S.\ DOE Office of Science User Facility operated by the Oak Ridge National Laboratory. 
\end{acknowledgments}

\newpage
\setcounter{equation}{0}
\setcounter{figure}{0}
\setcounter{table}{0}
\makeatletter
\renewcommand{\theequation}{S\arabic{equation}}
\renewcommand{\thefigure}{S\arabic{figure}}
\renewcommand{\bibnumfmt}[1]{[S#1]}
\renewcommand{\citenumfont}[1]{S#1}
\onecolumngrid
\begin{center}
	\textbf{{\large Supplemental Materials:\\Reduction of the ordered magnetic moment and its relationship to Kondo coherence in Ce$_{1-x}$La$_{x}$Cu$_{2}$Ge$_{2}$}}
\end{center}
\vspace{2ex}
\twocolumngrid

\begin{figure}[]
	\centering
	\includegraphics[width=1.0\linewidth]{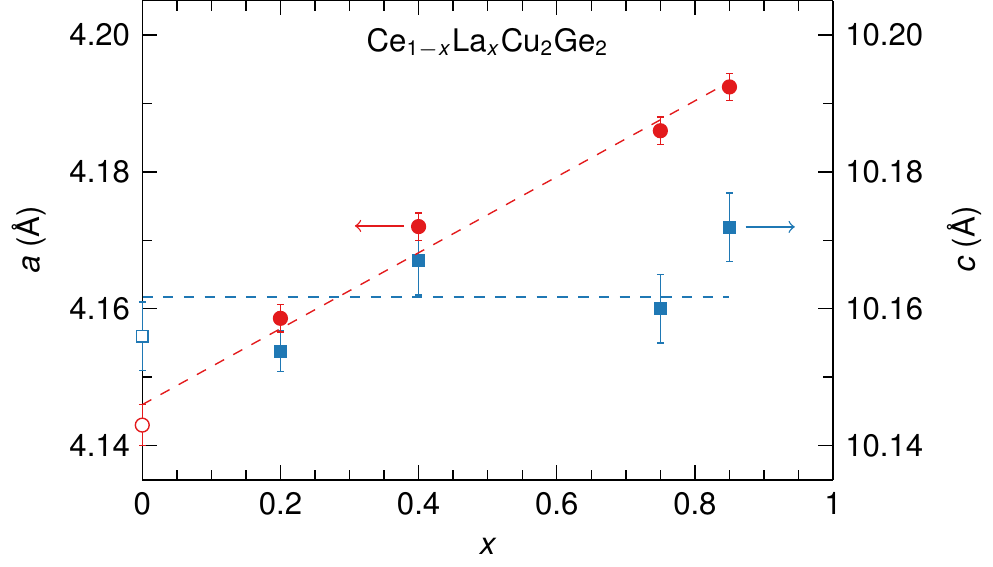}
	\caption{Lattice parameters for $x=0$, $0.2$. $0.40$, $0.75$, and $0.85$, at $T=1.5$, $1.6$, $0.3$, $0.3$, and $0.04$~K respectively.  Values for $x=0$ are from Ref.~\onlinecite{Krimmel_1997}. \label{FigS1}}
\end{figure}

\begin{figure}[]
	\centering
	\includegraphics[width=1.0\linewidth]{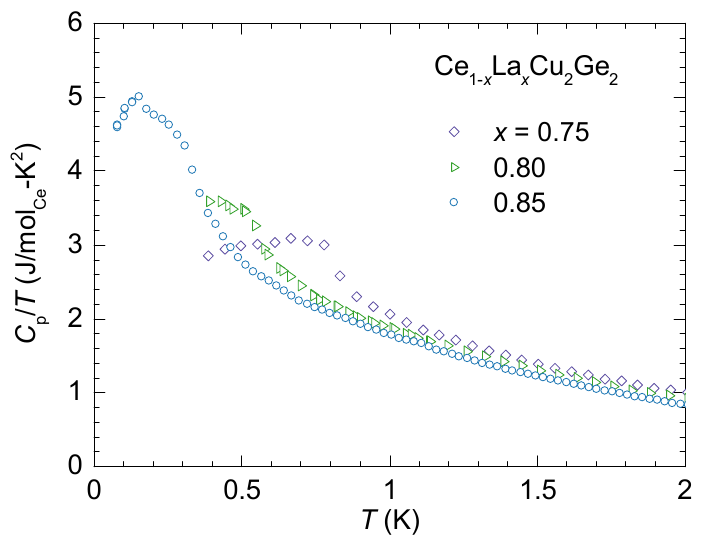}
	\caption{Heat capacity divided by temperature for $x=0.75$, $0.80$, and $0.85$.\label{FigS2}}
\end{figure}

Lattice parameters determined from fits to the $(1~1~0)$ and $(0~0~2)$ or $(0~0~4)$ structural Bragg peaks at the  temperatures listed in the figure caption are given in Fig.~\ref{FigS1}.  $a$ linearly increases by $1.2\%$ between $x=0$ and $0.85$.  $c$ may slightly increase with increasing $x$, but its perceived change is close to the resolution limit of the experiments.  The determined lattice parameters are consistent with previous ambient temperature data \cite{Hodovanets_2015}, and the replacement of Ce by larger La.

Figure~\ref{FigS2} shows the heat capacity divided by temperature $C/T$ in units of J$/$mol$_{\textrm{Ce}}$-K$^{2}$.  The plot illustrates that it is difficult to isolate the component of $C/T$ strictly due to the magnetic transition, because of the added complexity imposed by the contribution due to the Kondo effect.  A thorough analysis of $C(T)$ for these and other values of $x$ is given in Ref.~\onlinecite{Hodovanets_2015} and the associated Supplemental Material.

\end{document}